\documentclass[12pt]{article}
\pdfoutput=1
\usepackage{amsmath}
\usepackage{graphics,graphicx,xcolor}
\def\be{\begin{equation}}
\def\ee{\end{equation}}
\def\arr{\begin{array}{rll}}
\def\ea{\end{array}}
\def\bea{\begin{eqnarray}}
\def\eea{\end{eqnarray}}
\def\N2{$N{=}2$}

\def\>{\rangle}
\def\<{\langle}
\def\+{\dagger}
\def\={\ =\ }

\def\bal{\begin{aligned}}
\def\eal{\end{aligned}}
\begin{document}
\begin{titlepage}
\setcounter{page}{0}
\begin{center}
{\LARGE\bf  Perfect fluid equations with }\\
\vskip 0.5cm
{\LARGE\bf nonrelativistic conformal symmetry:}\\
\vskip 0.5cm
{\LARGE\bf Exact solutions }\\
\vskip 1.5cm
\textrm{\Large Anton Galajinsky \ }
\vskip 0.7cm
{\it
Tomsk Polytechnic University, 634050 Tomsk, Lenin Ave. 30, Russia} \\
\vskip 0.3cm
{\it
Tomsk State University of Control Systems and Radioelectronics,\\
Lenin ave. 40, 634050 Tomsk, Russia
} \\

\vskip 0.2cm
{e-mail: galajin@tpu.ru}
\vskip 0.5cm
\end{center}

\begin{abstract} \noindent
The group--theoretic approach is used to construct exact solutions to
perfect fluid equations invariant under the Schr\"odinger group, or
the $\ell$--conformal Galilei group, or the Lifshitz group. In each respective case,
the velocity vector field looks similar to the Bjorken flow. 
It is shown that 
one can reach an arbitrarily high density (and hence pressure) for a short period of time
by adjusting the value of $\ell$ and other free parameters available.
\end{abstract}

\vspace{0.5cm}

$\qquad$ Galilei group, the Lifshitz group
\end{titlepage}
\renewcommand{\thefootnote}{\arabic{footnote}}
\setcounter{footnote}0

\noindent
{\bf 1. Introduction}\\

Recent exploration of the fluid/gravity correspondence (for a review see \cite{MR}) 
stimulates a renewed interest in 
fluid mechanics with (non)relativistic conformal symmetry. The conventional formulation of fluid dynamics 
relies upon an expansion scheme in which
the effects of viscosity and heat transfer are implemented by means of specific
corrections to a perfect fluid stress-energy tensor. In the latter regard, symmetry 
analysis and exact solutions to (non)relativistic perfect fluid equations are of particular importance.

For a properly chosen equation of state, nonrelativistic
perfect fluid equations hold invariant under the Schr\"odinger group \cite{RS} (see also \cite{Ov}).\footnote{In this work, we are
primarily concerned with the nonrelativistic case. 
For a discussion of relativistic conformal fluids see e.g. \cite{FO} and references therein.}
As is well known, the Schr\"odinger algebra is a particular instance ($\ell=\frac 12$)
in the $\ell$-conformal Galilei algebra \cite{Henkel,NOR}. 
Transformations comprizing the $\ell$-conformal Galilei group include
(temporal) translation, dilatation, and special
conformal transformation, which form $SL(2,R)$ subgroup, 
as well as spatial rotations, spatial translations, Galilei boosts and a chain of 
constant accelerations. 
Structure relations of the corresponding Lie algebra involve an arbitrary (half)integer 
parameter $\ell$ (giving the name
to the group), 
which specifies the number of acceleration generators at hand \cite{Henkel,NOR} and 
ensures that the algebra is finite-dimensional.

Perfect fluid equations with the $\ell$-conformal Galilei symmetry have been formulated quite 
recently \cite{AG,AG1} (for related further developments see \cite{TS,TS1,TS2,TS3}).\footnote{Worth mentioning also is the
group-theoretic approach in \cite{BJLNP,NRR}.
The formalism is particularly suitable for taking into account constituent particles, 
which carry nonabelian charges or spin degrees of freedom, as well as for incorporating anomalies.} 
They include the continuity equation, which holds the conventional form, a generalization of the Euler equation
involving $2\ell$ material derivatives acting upon the velocity vector 
field,\footnote{Systems invariant under
the $\ell$-conformal Galilei group in general involve higher derivative terms (see e.g. \cite{GM4} and references therein).
The only known example without higher derivatives, in which all constants of motion 
are functionally independent, was built in \cite{CG} by making recourse 
to geodesics on Ricci-flat spacetimes with the $\ell$-conformal Galilei isometry
group.} and a properly chosen equation 
of state supporting the desired symmetry. Although realizations of the 
$\ell=\frac 12$ conformal Galilei group
in fluid mechanics have been extensively studied in the past 
(see e.g. \cite{RS,FO,HH2,JNPP,HH1,BMW,HZ} and references therein), exact solutions to
perfect fluid equations with the $\ell$-conformal Galilei symmetry have not yet 
been explored in any detail. The goal of this work is to fill in this gap.

Our strategy is to make use of the vast symmetry group 
characterizing the system and build exact solutions relying upon 
a properly chosen symmetry subgroup. 
As a technical tool, we use the group-theoretic approach in \cite{Ov,PO}. 
Within the framework of this method, one first constructs variables and fields invariant 
under the action of a specific subgroup of interest and then reduces the original 
complicated partial differential equations to simpler ones, which occasionally
may turn out to be ordinary differential equations.

If one disregards the
generator of special conformal transformation in the Schr\"odinger algebra 
($\ell=\frac 12$), the remaining structure
relations can be modified so as to include an arbitrary constant $z$ (known
as the dynamical critical exponent), thus giving rise to the Lifshitz algebra. The
Lifshitz holography attracted recently considerable attention 
(for a review see \cite{MT}). The group-theoretic construction of solutions to 
perfect fluid equations with the $\ell$-conformal Galilei symmetry can be immediately
extended to the case of the Lifshitz group. One of the objectives of this work
is to build exact solutions to perfect fluid equations with the 
Lifshitz symmetry group, which have been recently formulated in \cite{AG1}. 

The work is organized as follows. In the next section, basic facts concerning
fluid mechanics with the $\ell$-conformal Galilei symmetry are briefly reminded.

In Sect. 3.1, exact solutions to perfect fluid equations with the 
$\ell$-conformal Galilei symmetry in $1+1$ dimensions are constructed. 
Within the framework of the group-theoretic approach \cite{Ov,PO}, each one-dimensional subgroup 
is analyzed in turn. 
It is shown that the most interesting specimen in the family
of exact solutions is associated with the subgroup of scaling transformations. 
In particular, for $d=1$ and $\ell=\frac 12$ the analysis can be accomplished in 
full generality.
The scaling-invariant variables and fields are built in terms of which
the continuity equation and the Euler equation reduce to 
ordinary differential equations. In $1+1$ dimensions, two possibilities are available to resolve the latter.
Either one can use the continuity equation to fix the density and the Euler equation
to determine the velocity or vice versa. The latter option, 
in which the continuity equation and the Euler equation effectively swap over 
their roles, looks rather unusual. Yet, for $\ell> \frac 12$ and arbitrary spatial dimension $d$ 
it proves to be the only reliable way. Interestingly enough, the resulting 
fluid velocity $\upsilon (t,x)=\frac{\ell x}{t}$, where $t$ and $x$ stand for the
temporal and spatial coordinates, differs from the celebrated 
Bjorken flow \cite{B} by a constant factor $\ell$ only and coincides with it at $\ell=1$. As 
a result, for greater values of $\ell$ a fluid moves faster.
The dependence of density upon $\ell$ is more subtle and is discussed below.

In Sect. 3.2, the analysis is extended to arbitrary spatial 
dimension $d$. In particular, an exact solution to perfect fluid equations with the 
$\ell$-conformal Galilei symmetry is built which links to the 
subgroup of scaling transformations. The resulting velocity vector 
field $\upsilon_i (t,x)=\frac{\ell x_i}{t}$, where $t$ and $x_i$, with $i=1,\dots,d$, denote
temporal and spatial coordinates, is a natural 
generalization of the Bjorken flow \cite{B} to higher dimensions. Qualitative behaviour of a fluid moving in $1+d$ dimensions is similar 
to the $1+1$ dimensional case. The greater the value of $\ell$, the faster a fluid moves. 
It is also demonstrated 
that 
one can reach an arbitrarily high density (and hence
pressure) for a short period of time by adjusting the value of $\ell$ 
and other free parameters available. This allows one to suggest that 
fluid equations with the $\ell$-conformal Galilei symmetry may prove 
useful in other physical contexts such as quark-gluon plasma, cosmology of the early universe
and 
physics of explosion phenomena. Other interesting solutions, which follow  
by applying special conformal transformation and constant acceleration transformation,
are also discussed.

In Sect. 4, we carry out a similar analysis for perfect fluid equations with the 
Lifshitz symmetry group \cite{AG1}. Firstly, some basic facts concerning
fluid mechanics with the Lifshitz symmetry are briefly reminded. Secondly,
a particular solution is built 
which links to the subgroup of anisotropic scaling transformations.  
In general, the greater the value of $z$, the slower a fluid moves,
$\upsilon_i (t,x)=\frac{x_i}{z t}$ being the corresponding velocity vector field. 
Note that, as compared to 
the $\ell$-conformal Galilei group, the Euler equation with the Lifshitz symmetry 
involves only one material derivative.  Our analysis also points at the lower bound $z> \frac 12$
for the dynamical critical exponent. Interestingly enough, a similar bound has recently 
been revealed in \cite{AG4}, when building 
dynamical realizations of the Lifshitz group in mechanics and general relativity.
 
In Sect. 5, we briefly discuss how the analysis in the preceding sections can be extended
to the case of a viscous fluid with the $\ell$-conformal Galilei symmetry.

We summarize our results and discuss possible further developments in the concluding Sect. 6.

Throughout the paper, summation over repeated indices is understood unless otherwise stated.

\vspace{0.5cm}

\noindent
{\bf 2. Perfect fluid equations with the $\ell$-conformal Galilei symmetry}\\

\noindent
The action of the $\ell$-conformal Galilei group in
a nonrelativistic spacetime parametrized by a temporal 
coordinate $t$ and spatial variables $x_i$, $i=1,\dots,d$, reads 
(no summation over repeated index $n$ in the last formula)
\cite{Henkel,NOR}
\begin{align}\label{tr}
&
t'=\frac{\alpha t+\beta}{\gamma t+\delta}, &&
x'_i={\left(\frac{\partial t'}{\partial t} \right)}^\ell x_i;
\nonumber\\[2pt]
&
t'=t, && x'_i=x_i+a^{(n)}_i t^n,
\end{align}
where $(\alpha,\beta,\gamma,\delta)$ obeying the constraint $\alpha \delta-\beta \gamma=1$
parametrize $SL(2,R)$-subgroup, and $a^{(n)}_i$, in which 
$n=0,\dots, 2\ell$ and
$\ell$ is a (half)integer real number, 
describe spatial translation ($n=0$), the Galilei boost ($n=1$), 
and higher order constant accelerations ($n=2,\dots,2\ell$). 
The group also involves $SO(d)$-rotation acting upon $x_i$, 
which in what follows will be disregarded. 

From (\ref{tr}) one finds generators of infinitesimal (temporal) translation, 
dilatation, special conformal transformation, and accelerations (for more details 
see e.g. \cite{AG})
\be\label{gen}
H=\frac{\partial}{\partial t}, \qquad D=t \frac{\partial}{\partial t}+\ell x_i \frac{\partial}{\partial x_i}, \qquad K=t^2 \frac{\partial}{\partial t}+2 \ell t x_i \frac{\partial}{\partial x_i}, 
\qquad C^{(n)}_i=t^n  \frac{\partial}{\partial x_i},
\ee
which obey the structure relations the $\ell$-conformal Galilei algebra \cite{Henkel,NOR}
\begin{align}\label{algebra}
&
[H,D]=H, &&  [H,K]=2 D, && [D,K]=K,
\nonumber\\[2pt]
&
[H,C^{(n)}_i]=n C^{(n-1)}_i, && [D,C^{(n)}_i]=(n-l) C^{(n)}_i, && [K,C^{(n)}_i]=(n-2l) C^{(n+1)}_i.
\end{align}
Note that the last commutator constraints 
$\ell$ to be a (half)integer number, which is needed in order to ensure that
the algebra is
finite-dimensional.

Let $\rho(t,x)$ and $\upsilon_i (t,x)$, $i=1,\dots,d$, be
the density and the velocity vector field of a perfect fluid. 
Transformation law of $\rho(t,x)$ under (\ref{tr})
is obtained by fixing a value of the temporal variable 
$t$ and demanding the mass of a $d$-dimensional volume 
element to be invariant. 
This yields
\be\label{trr}
\rho(t,x)= {\left(\frac{\partial t'}{\partial t} \right)}^{\ell d} \rho' (t',x'),
\ee
for the $SL(2,R)$-transformation and
\be\label{trr1}
\rho(t,x)=\rho' (t',x'),
\ee
for the accelerations.

Considering an orbit of a liquid particle parametrized by $x_i (t)$, from 
the equation $\frac{d x_i (t)}{d t}= \upsilon_i (t,x(t))$ and 
the transformation (\ref{tr}) one obtains\footnote{When restricting 
the transformation (\ref{tr})
to a particle orbit, one replaces $x'_i$ with $x'_i (t')$ and 
$x_i$ with $x_i (t)$.} 
\be\label{trv}
\upsilon_i (t,x)={\left(\frac{\partial t'}{\partial t} \right)}^{1-\ell} \upsilon'_i (t',x')+\frac{\partial}{\partial t} {\left(\frac{\partial t'}{\partial t} \right)}^{-\ell} x'_i,
\ee
for the $SL(2,R)$-transformation and (no sum over repeated index $n$)
\be\label{trv1}
\upsilon_i (t,x)=\upsilon'_i (t',x')-n a^{(n)}_i t^{n-1},
\ee
for the accelerations.

Perfect fluid equations of motion, which hold invariant under the action of the 
$\ell$-conformal Galilei group, were formulated in \cite{AG}
\bea\label{pfl}
&&
\frac{\partial \rho}{\partial t} + 
\frac{\partial ( \rho \upsilon_i )}{\partial x_i}=0, \qquad
\rho  \mathcal{D}^{2\ell} \upsilon_i=-\frac{\partial p}{\partial x_i},
\qquad
p=a \rho^{1+\frac{1}{\ell d}},
\eea
where $\mathcal{D}$ is 
the material derivative
\be\label{MD}
\mathcal{D}=\frac{\partial}{\partial t} +
\upsilon_i  (t,x) \frac{\partial}{\partial x_i}
\ee
and $a$ is a positive constant. As is seen from (\ref{pfl}), 
the continuity equation maintains its conventional form, whereas 
the Euler equation 
involves $2\ell$ 
material derivatives acting upon the velocity vector field. The equation of state, 
which links the pressure $p$ to the
density $\rho$ in the last equation entering (\ref{pfl}), is chosen so as to provide the invariance of the Euler equation.
It should be mentioned that the instance of $\ell=\frac 12$ was 
thoroughly investigated in \cite{RS,JNPP} (see also the discussion in \cite{HZ}).

Interestingly enough, transformation laws of 
$\mathcal{D}^{n} \upsilon_i$ with $n>1$ under the $SL(2,R)$-subgroup
defined in (\ref{tr}) and (\ref{trv}) prove to involve higher order Schwarzian 
derivatives \cite{AG3}. Fluid mechanics with the $\ell$-conformal Galilei symmetry
appears to be the first physical context which naturally incorporates the higher order Schwarzians.

\vspace{0.5cm}

\noindent
{\bf 3. Perfect fluid equations with the $\ell$-conformal Galilei symmetry: Exact solutions}\\

\vspace{0.2cm}
\noindent
{\it 3.1. Exact solutions in $1+1$ dimensions}\\

We proceed to constructing exact solutions to the perfect fluid equations with the $\ell$-conformal Galilei symmetry 
by first analyzing the case of
$d=1$ and $\ell=\frac 12$. This simpler setting helps us to grasp essential 
features which will pertain to a more realistic case of arbitrary $d$ and $\ell$.

As was mentioned in the Introduction, our strategy is to make use of the 
vast symmetry group 
characterizing the system and to build explicit solutions relying upon a properly chosen symmetry subgroup. 
As a technical tool, one can use the group-theoretic approach \cite{Ov,PO}, 
which first constructs variables and fields invariant 
under the action of a specific subgroup of interest and then reduces the original 
complicated partial differential equations to simpler ones, which occasionally
may turn out to be ordinary differential equations.

Given the  $\ell=\frac 12$ conformal Galilei group in $1+1$ dimensions, let us first focus on
the one-dimensional subgroup of scaling transformations. The corresponding generator $D$ is obtained from
eqs. (\ref{tr}), (\ref{trr}), (\ref{trv}) by choosing 
$\alpha=e^{\frac{\lambda}{2}}$, $\delta=e^{-\frac{\lambda}{2}}$, $\beta=0$, $\gamma=0$ in (\ref{tr}), 
setting $\lambda$ to be an infinitesimal parameter and Taylor expanding in $\lambda$ 
up to the first order\footnote{Here and in what follows $D$ used to designate the dilatation 
generator is not to be confused with the calligraphic
$\mathcal{D}$ reserved for the material derivative (\ref{MD}).}
\be\label{dil}
D=t \partial_t+\frac 12 x \partial_x-\frac 12 \rho \partial_{\rho}
-\frac 12 \upsilon \partial_{\upsilon},
\ee 
where we abbreviated $\partial_t=\frac{\partial}{\partial t}$ etc.

As the next step, one builds variables and fields which hold invariant under the scaling transformation 
by switching to the characteristic equations\footnote{For a detailed account of 
the method of characteristics see e.g. \cite{SM}.} 
\be\label{char}
\frac{dt}{t}=\frac{2 d x}{x}=-\frac{2 d \rho}{\rho}=-\frac{2 d \upsilon}{\upsilon}
\ee
associated with the linear partial differential 
equation $D f(t,x,\rho,\upsilon)=0$ with $D$ in (\ref{dil}) and unknown $f(t,x,\rho,\upsilon)$.
Three first integrals of (\ref{char})
\be
\frac{x^2}{t}=C_1, \qquad x \upsilon=C_2, \qquad x \rho=C_3,
\ee
allow one to build the 
scale-invariant variable
\be
\frac{x^2}{t}:=y
\ee
and two scale-invariant functions $u(y)$ and $w(y)$ 
\be\label{rv}
x \upsilon(t,x):=u(y), \qquad x \rho(t,x):=w(y).
\ee
Because one is primarily concerned with $\rho(t,x)$ and $\upsilon(t,x)$, 
the point $x=0$ should be 
excluded from the consideration which also eliminates $y=0$.
In what follows, we also assume the condition $t>0$ to hold.

Being rewritten in terms of the scale-invariant objects, the continuity equation and 
the Euler equation reduce to the ordinary differential equations
\be\label{2eq}
\frac{d}{d y} \left(\left(u-\frac{y}{2} \right) \frac{w}{y} \right)=0, \qquad 
\frac{d}{d y} \left(\frac{u^2+3 a w^2}{y}-u \right)=0,
\ee
where $a$ is a constant which enters the equation of state $p=a \rho^3$,
two first integrals of which allow one to find $u$ and $w$ algebraically.

Two options are available at this point. If $u\ne \frac{y}{2}$, the leftmost (continuity) equation 
gives $w(y)\sim\frac{y}{u-\frac{y}{2}}$, while
the rightmost (Euler) equation yields a quartic algebraic equation to determine $u(y)$
\be\label{seq}
u(y)=\frac{y}{2}\pm \frac 12 \sqrt{ {\left(\frac{y}{2}+c_1\right)}^2-c_1^2 }
\pm \frac 12 \sqrt{{\left(\frac{y}{2}+c_2\right)}^2-c_2^2}, \qquad 
w(y)=\left( \frac{c_1-c_2}{4 \sqrt{3 a}}\right) \frac{y}{u-\frac{y}{2}},
\ee
where $c_1$, $c_2$ are constants of integration. For physical reasons, the signs of $c_1$, $c_2$,
$\pm$ signs in $u(y)$, and the domain $x>0$ or $x<0$ should altogether be balanced so as 
to keep the density $\rho(t,x)=\frac{w(y)}{x}$ positive. 

Alternatively, one can resolve the continuity equation in (\ref{2eq}) by setting 
\be
u(y)=\frac{y}{2}
\ee
and then use the Euler equation to fix the form of $w(y)$
\be
w(y) =\pm \frac{1}{\sqrt{3 a}} \sqrt{ {\left(\frac{y}{2}+c \right)}^2-c^2 },
\ee
where $c$ is a constant of integration. As above, $\pm$ sign in $w(y)$ and the domain
$x>0$ or $x<0$ should be chosen in such a way that the density $\rho(t,x)=\frac{w(y)}{x}$ is positive.

To summarize, for $d=1$ and $\ell=\frac 12$ we have constructed six solutions which link to
the one-dimensional subgroup of scaling transformations in the $\ell$-conformal Galilei group.

For an arbitrary value of $\ell$ and $d=1$, one can proceed along similar lines. 
The generator of infinitesimal scaling transformation reads
\be\label{dil1}
D=t \partial_t+\ell x \partial_x-\ell \rho \partial_{\rho}
-(1-\ell) \upsilon \partial_{\upsilon},
\ee 
which gives rise to the scale-invariant variable 
\be
\frac{x^2}{t^{2\ell}}:=y, 
\ee
and the scale-invariant fields
\be\label{rv1} 
x^{\frac{1-\ell}{\ell}} v(t,x):=u(y), \qquad x  \rho(t,x):=w(y).
\ee
The continuity equation can then be cast into the form
\be 
\frac{d}{d y} \left(\left(u-\ell y^{\frac{1}{2\ell}} \right) 
\frac{w}{y^{\frac{1}{2\ell}}} \right)=0,
\ee
which again implies two possibilities: either $u=\ell y^{\frac{1}{2\ell}}$ or $u\ne \ell y^{\frac{1}{2\ell}}$.

Unfortunately, in the latter case it proves problematic to integrate the Euler equation even 
for the simplest case of $\ell=1$.
So in what follows we focus on the particular solution 
\be\label{v}
u=\ell y^{\frac{1}{2\ell}} \quad \Rightarrow \quad v(t,x)=\frac{\ell  x}{t},
\ee
with $t>0$. Interestingly enough, the resulting 
 velocity in (\ref{v}) differs from the celebrated 
Bjorken flow \cite{B} by a constant factor only and coincides with it at $\ell=1$.

Given $v(t,x)$ in (\ref{v}), the analysis of the Euler equation proceeds differently for integer and 
half integer values of the parameter $\ell$. Taking into account the identity
\be
\mathcal{D}^{2\ell} \left(\frac{\ell  x}{t}\right)=\ell (\ell-1)(\ell-2)\dots (\ell-2\ell) \frac{x}{t^{2\ell+1}}, 
\ee
where $\mathcal{D}$ is the material derivative (\ref{MD}),
one concludes that $\mathcal{D}^{2\ell} v(t,x)=0$ for integer values of $\ell$ meaning that the pressure 
(and hence the density) depends on the temporal variable only, i.e. a fluid is homogeneous. The constraint 
$\partial_x \rho(t,x)=0$ along with 
the definition of $w(y)$ in (\ref{rv1}) then yield
\be\label{rr2}
2 y w'(y)-w(y)=0 \qquad \Rightarrow \qquad w(y)=c \sqrt{y} \qquad \Rightarrow \qquad
\rho(t,x)=\frac{c}{t^\ell},
\ee
where $c>0$ is a constant of integration. 

For half-integer $\ell$, one can cast the Euler equation into the form 
\be
\frac{d}{d y} \left(\frac{w^{\frac{1}{\ell}}}{y^{\frac{1}{2\ell}}}+\frac{\ell (\ell-1)(\ell-2)\dots 
(\ell-2\ell) y}{2 a (\ell  +1)} \right)=0,
\ee 
where $a$ is a positive constant entering the equation of state 
$p=a \rho^{1+\frac{1}{\ell}}$,
which is easily integrated and via (\ref{rv1}) it determines the density
\be\label{r32}
\rho(t,x)= {\left(\frac{c}{t}-\frac{\ell (\ell-1)(\ell-2)\dots 
(\ell-2\ell) x^2}{2 a (\ell  +1) t^{2\ell+1}} \right)}^\ell,
\ee
where $c>0$ is a constant of integration and $t>0$. Note that the latter 
relation is formally valid for
integer $\ell$ as well. Indeed, the second term in braces vanishes for integer $\ell$ and the 
formula reduces to (\ref{rr2}).

Because the density and pressure are assumed 
to be positive-definite functions and the constant $\ell (\ell-1)(\ell-2)\dots 
(\ell-2\ell)$ alternates in sign as $\ell$ increases, one is ultimately forced to choose
\be\label{int}
\ell=\frac{1+4k}{2}
\ee
with $k=0,1,2,\dots$, which guarantees that the expression under the square root in (\ref{r32}) is positive-definite.

%======================= Fig.1===========================>
\begin{figure}[ht]
\begin{center}
\resizebox{0.5\textwidth}{!}{%
\includegraphics{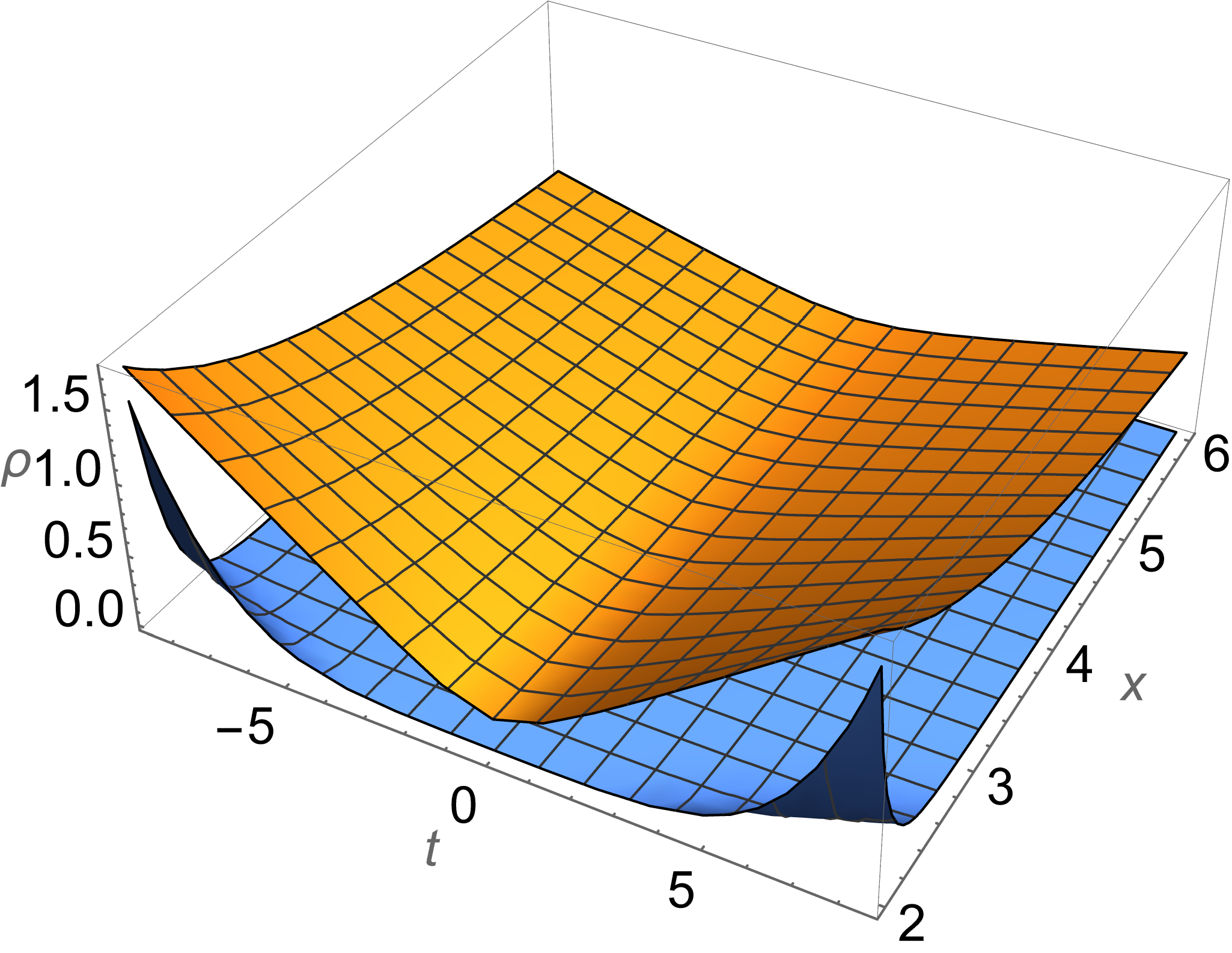}}\vskip-4mm
\caption{\small The graph of the density $\rho$ as the function of $t$ and $x$ for $\ell=\frac 12$ (orange) and $\ell=\frac 52$ (blue) 
with 
$t \in [2,6]$, $x \in [-8.5,8.5]$, $c=0.1$, $a=0.5$.}
\label{fig1}
\end{center}
\end{figure}

As is seen from (\ref{v}), for greater values of $\ell$ a fluid 
always moves faster. The dependence of density upon $\ell$ is more subtle, however. 
For integer $l$,
fluids with greater $\ell$ are denser on the time interval $0<t<1$, whereas for $t>1$ the smaller 
$\ell$ the denser a fluid (see (\ref{rr2}) above). For half-integer $\ell$ belonging to the sequence (\ref{int}), 
the qualitative behaviour of $\rho_\ell (t,x)$ is similar. Given $\ell_1 > \ell_2$ from (\ref{int}), 
$\rho_{\ell_1} > \rho_{\ell_2}$ on a certain time interval $0<t<\tilde t$, whereas 
$\rho_{\ell_1} < \rho_{\ell_2}$ for $t>\tilde t$. In each respective case, the higher the value of $\ell$, the sharper 
the downfall of density in the vicinity of certain threshold value of the temporal variable $t$. Fig. 1 depicts surfaces $\rho=\rho(t,x)$ 
for $\ell=\frac 12$ (orange) and $\ell=\frac 52$ (blue) with
$t \in [2,6]$, $x \in [-8.5,8.5]$, $c=0.1$, $a=0.5$.

Having constructed solutions associated with the subgroup of scaling transformations
in the $\ell$-conformal Galilei group, one can analyze
other one-dimensional subgroups in a similar fashion. Let us discuss them in turn.

For the subgroup generated by $H=\partial_t$ 
the invariant variable is $x$. As far as dynamics is concerned, the corresponding 
stationary solutions are not of particular interest.

Given a (half)integer $\ell$, there are $2\ell+1$ one-dimensional subgroups 
associated with the acceleration generators
\be
C^{(n)}=t^n \partial_x+n t^{n-1} \partial_v, 
\ee 
where $n=0,1,\dots,2\ell$. In this case, the invariant variable is $t$ and the invariant 
fields read
\be\label{if3}
\rho(t), \qquad t v(t,x)-n x:=t u(t), 
\ee
Note that the density depends of the temporal variable only. The continuity equation then yields
\be
\rho(t)=\frac{c}{t^n}, 
\ee
where $c>0$ is a constant of integration and $t>0$, while analysis of 
the Euler equation reveals that $u(t)$ in (\ref{if3}) is the rational function of the form
\be\label{seq1}
u(t)=\frac{c_{-1} (n)}{t}+\sum_{k=0}^{2\ell-1} c_k (n) t^k,
\ee
where the number coefficients $c_{-1} (n)$ and $c_k (n)$ depend on specific $\ell$ and $n$ chosen.
For physical reasons, 
the vast majority of such solutions should be discarded as the corresponding velocity 
increases unbounded with time (runaway solutions).
The only viable variant reads
\be
\rho(t)=\frac{c}{t^n}, \qquad v(t,x)=\frac{n x+c_{-1}(n)+c_0 (n) t}{t},
\ee
where constants $c_{-1}(n)$ and $c_0 (n)$ depend on specific $\ell$ and $n$ chosen and 
should be fixed directly from the Euler equation. Notice again that given $\ell$ there are $2\ell+1$ such solutions
which correspond to $n=0,1,\dots,2\ell$.

Finally, the subgroup of special conformal transformations associated with
\be
t'=\frac{t}{1-\gamma t}
\ee
in (\ref{tr}), (\ref{trr}), (\ref{trv}) is generated by
\be
K=t^2 \partial_t+2\ell t x \partial_x-2 t \ell \rho \partial_\rho+
\left(2 \ell x+2 t (\ell-1) \upsilon \right) \partial_\upsilon.
\ee
Although the invariant variable and one invariant field  are easily constructed
\be
\frac{x}{t^{2\ell}}:=y, \qquad x \rho(t,x):=w(y),
\ee
it proves problematic to completely separate variables in the associated system of characteristic equations (see e.g. \cite{SM}))
and thus introduce an invariant counterpart $u(y)$ for the 
velocity $v(t,x)$. So this subgroup appears to be of little help for constructing 
explicit solutions.

To summarize, our analysis above reveals that the most interesting specimen in the family of exact 
solutions to the
perfect fluid equations with the $\ell$-conformal Galilei symmetry in $1+1$ dimensions 
is the one associated with the subgroup of scaling transformations. In the next section, 
we construct
its analogue in arbitrary spatial dimension.

\vspace{0.5cm}

\noindent
{\it 3.2. Exact solutions in arbitrary dimension}\\

The results in the preceding section can be generalized 
to the case of arbitrary spatial dimension, although the ensuing dynamical equations are no longer
ordinary differential equations. Like above, we will be primarily concerned 
with solutions which link to the subgroup of scaling transformations generated by
\be\label{dil2}
D=t \partial_t+\ell x_i \partial_{x_i}-\ell d \rho \partial_{\rho}
-(1-\ell) \upsilon_i \partial_{\upsilon_i},
\ee 
where $d$ is the spatial dimension and $i=1,\dots,d$. From (\ref{dil2})
one finds the scale-invariant variables and fields
\be\label{vaf}
\frac{x_i}{t^\ell}:=y_i, \qquad t^{\ell d} \rho(t,x):=w(y), 
\qquad t^{1-\ell} \upsilon_i (t,x):=u_i (y),
\ee
with $t>0$.

%======================= Fig.2===========================>
\begin{figure}[ht]
\begin{center}
\resizebox{0.4\textwidth}{!}{%
\includegraphics{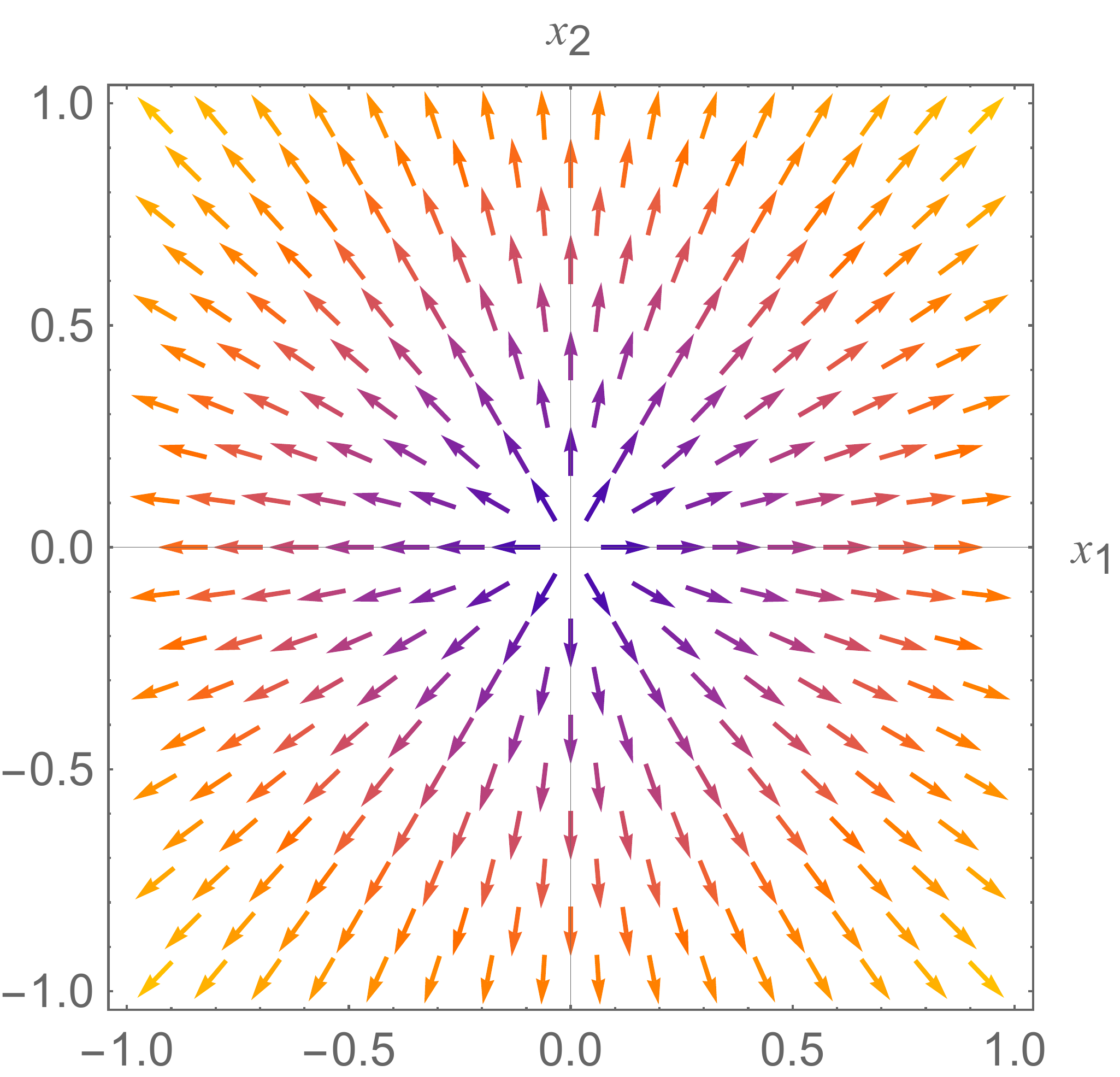}}\vskip-4mm
\caption{\small A flow generated by the vector field 
$\upsilon_i=\frac{\ell x_i}{t}$ in two spatial dimensions ($i=1,2$) for $x_1 \in [-1,1]$, $x_2 \in [-1,1]$ at $\ell=1$ and $t=10$.
}
\label{fig2}
\end{center}
\end{figure}

The key observation is that the continuity equation is drastically simplified when 
rewritten in terms of the invariant objects
\be
\frac{\partial}{\partial y_i} \left(w \left(u_i-\ell y_i \right) \right)=0. 
\ee
Instead of solving this partial differential equation in full generality, 
we choose a simpler road and use it to fix 
the velocity vector field
\be\label{VVF}
u_i=\ell y_i \qquad \Rightarrow \qquad \upsilon_i=\frac{\ell x_i}{t}, 
\ee
where $t>0$, which is a natural generalization of the Bjorken flow to arbitrary spatial dimension. For such a choice, one readily obtains
\be
\mathcal{D}^{2\ell} \upsilon_i=(\ell-1)(\ell-2)\dots (\ell-2\ell) \frac{\upsilon_i}{t^{2\ell}},
\ee 
which naturally separates integer
and half-integer values of the parameter $\ell$.  
Fig. 2 depicts a flow generated by
the vector field 
(\ref{VVF}) in two spatial dimensions ($i=1,2$) for $x_1 \in [-1,1]$, $x_2 \in [-1,1]$ 
at $\ell=1$ and $t=10$.

For integer $\ell$, the previous formula gives $\mathcal{D}^{2\ell} \upsilon_i=0$ 
and the Euler equation simplifies to 
$\partial_i p=0$, where $p=a \rho^{1+\frac{1}{\ell d}}$ is the pressure,
meaning that the density depends on the 
temporal variable only (i.e. the fluid is homogeneous)
\be\label{rrr33}
\rho(t,x)=\frac{c}{t^{\ell d}},
\ee
where $c>0$ is a constant of integration. In obtaining the latter relation, we differentiated the second equation in (\ref{vaf}) with respect to
$x_i$, which gives $\frac{\partial w(y)}{\partial y_i}=0 ~ \Rightarrow ~ w(y)=c$.

For half-integer $\ell$, the Euler equation can be cast into the form
\be\label{den}
\frac{\partial}{\partial y_i} \left(w^{\frac{1}{\ell d}}+ \frac{\ell (\ell-1)(\ell-2)\dots (\ell-2\ell) y_j y_j }{2 a (1+\ell d)} 
 \right)=0,
\ee
where $a>0$ is the parameter entering the equation of state $p=a \rho^{1+\frac{1}{\ell d}}$,
which can be easily solved
\be
w(y)={\left(c-\frac{\ell (\ell-1)(\ell-2)\dots (\ell-2\ell) y_i y_i}{2 a (1+\ell d)} 
 \right)}^{\ell d},
\ee
where $c>0$ is a constant of integration.  The density scalar field is then specified by the second equation in (\ref{vaf})
\be\label{r33}
\rho(t,x)=
{ \left( \frac{c}{t}
-\frac{\ell (\ell-1)(\ell-2)\dots (\ell-2\ell) x_i x_i }{2 a (1+\ell d) t^{2\ell+1}} \right) }^{\ell d},
\ee
where $a$ is a positive constant entering the equation of state 
$p=a \rho^{1+\frac{1}{\ell d}}$.
Notice that the resulting fluid is isotropic 
and (\ref{r33}) is formally valid for
integer $\ell$ as well. Indeed, the second term in braces vanishes for integer $\ell$ and the 
formula reduces to (\ref{rrr33}).

Because the density and the pressure are assumed 
to be positive-definite functions and the constant 
$\ell (\ell-1)(\ell-2)\dots (\ell-2\ell)$ alternates in sign as $\ell$ 
increases, one is ultimately forced to choose
\be\label{SEq1}
\ell=\frac{1+4k}{2}
\ee
with $k=0,1,2,\dots$, which guarantees that the expression under the square root in (\ref{r33}) is positive-definite.

To summarize, in arbitrary spatial dimension a particular solution to the perfect fluid equations 
with the $\ell$-conformal Galilei symmetry, which is associated with the subgroup
of scaling transformations, reads
\be\label{SIS}
\upsilon_i=\frac{\ell x_i}{t}, \qquad \rho(t,x)=
{ \left( \frac{c}{t}
-\frac{\ell (\ell-1)(\ell-2)\dots (\ell-2\ell) x_i x_i }{2 a (1+\ell d) t^{2\ell+1}} \right) }^{\ell d},
\ee
where $\ell$ is either integer or half-integer belonging to the sequence $\ell=\frac{1+4k}{2}$, with $k=0,1,2,\dots$,
$a>0$ is a constant which contributes to the equation of state 
$p=a \rho^{1+\frac{1}{\ell d}}$, $c$ is an arbitrary positive constant and $t>0$.

%======================= Fig.3===========================>
\begin{figure}[ht]
\begin{center}
\resizebox{0.5\textwidth}{!}{%
\includegraphics{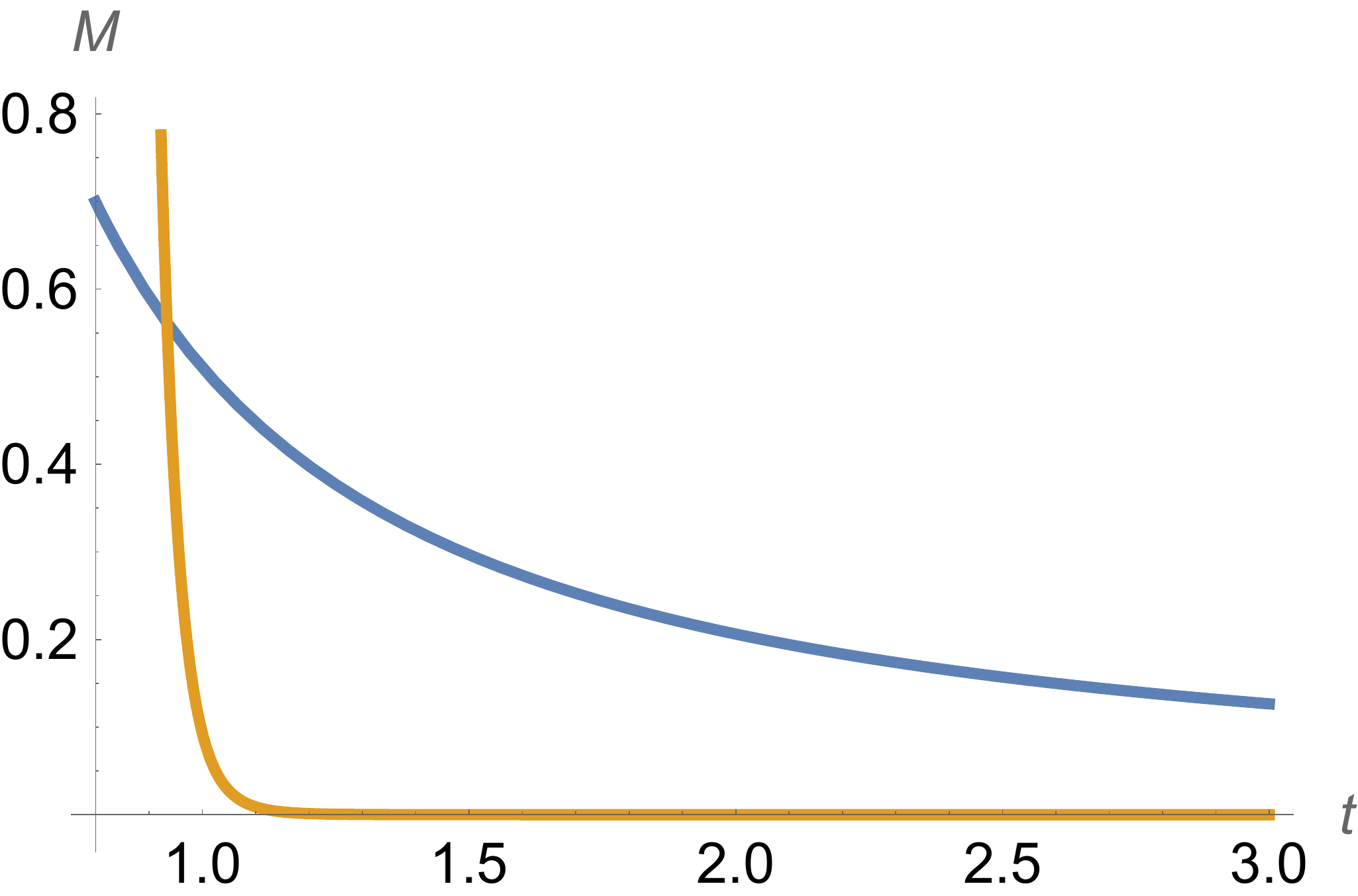}}\vskip-4mm
\caption{\small The dependence of the mass $M$ of a disk of unit radius 
centered at the origin of the coordinate system upon time $t$ 
for $\ell=\frac 12$ (blue)
and $\ell=\frac 52$ (orange) in two spatial dimensions with $c=0.1$, $a=0.5$, and $t \in [0.8,3]$.
}
\label{fig3}
\end{center}
\end{figure}

As is evident from (\ref{SIS}), a fluid moves faster
for greater values of $\ell$. 
For integer $l$,
fluids with greater $\ell$ are denser on the time interval $0<t<1$, whereas 
for $t>1$ the smaller $\ell$ the denser a fluid.
For half-integer $\ell$ belonging to the sequence (\ref{SEq1}), 
the dependence of density upon $\ell$ is more subtle. In general, it correlates with
the values of the free parameters $c$ and $a$ chosen. For example, Fig. 3 depicts
the dependence of the mass $M$ of a disk of unit radius 
centered at the origin of the coordinate system upon time $t$ for $\ell=\frac 12$ (blue)
and $\ell=\frac 52$ (orange) in two spatial dimensions with $c=0.1$, $a=0.5$, and $t \in [0.8,3]$.
 
Recall that, given a vector field $\upsilon_i$, the tensors 
\be
\frac{\partial \upsilon_i}{\partial x_j} 
-\frac{\partial \upsilon_j}{\partial x_i}, \qquad
\frac{\partial \upsilon_i}{\partial x_j} 
+\frac{\partial \upsilon_j}{\partial x_i}-\frac{2}{d} \delta_{ij} 
\frac{\partial \upsilon_k}{\partial x_k}, \qquad
\delta_{ij}
\frac{\partial \upsilon_k}{\partial x_k},
\ee
determine its vorticity, shear, and expansion. For $\upsilon_i$ in (\ref{SIS}), the first two
tensors vanish, while the latter one reduces to $\delta_{ij} \frac{d \ell}{t}$. Thus,
within the context of fluid mechanics, the group parameter $\ell$ links to 
the rate of expansion.

The fact that one can reach arbitrarily high density 
(and hence pressure) for a short period of time by adjusting the value of $\ell$ (and 
other free parameters at hand\footnote{The density in (\ref{SIS}) involves three adjustable 
parameters $\ell$, $c$, and $a$. Because the equations of motion (\ref{pfl}) hold
invariant under temporal translation, one more free parameter $t_0$ can be introduced 
by the obvious 
substitution $t\to t+t_0$ in (\ref{SIS}).})
allows one to presume that the fluid equations with the $\ell$-conformal Galilei symmetry
(\ref{pfl}) may prove
useful in other physical contexts such as the quark-gluon plasma, cosmology of the early universe and 
physics of explosion phenomena.

Having constructed a particular solution to (\ref{pfl}), which links to the subgroup of scaling transformations, 
one can build other interesting solutions by
applying to it symmetry transformations available. For example, the action of the 
special conformal transformation in nonrelativistic spacetime reads
\be
t'=\frac{t}{1-\gamma t}, \qquad
x'_i={\left(\frac{\partial t'}{\partial t} \right)}^\ell x_i,
\ee
where $\gamma$ is a (finite) parameter of the dimension ${[t]}^{-1}$.
Given a particular solution $\rho(t,x)$, $\upsilon_i (t,x)$ to the equations of motion, its one-parameter deformation
associated with the special conformal transformation reads
\bea
&&
\rho'(t,x)={\left(1+\gamma t \right)}^{-2 \ell d} 
\rho\left(\frac{t}{1+\gamma t},{\left(1+\gamma t \right)}^{-2 \ell} x \right),
\nonumber\\[2pt]
&&
\upsilon'_i (t,x)={\left(1+\gamma t \right)}^{2(\ell-1)} 
\upsilon_i \left(\frac{t}{1+\gamma t},{\left(1+\gamma t \right)}^{-2 \ell} x \right)+
\frac{2\gamma \ell x_i}{1+\gamma t}.
\eea
Being applied to (\ref{SIS}), this gives
\be
\upsilon'_i (t,x)=\frac{\ell\left(1+2\gamma t \right) x_i}{t\left(1+\gamma t \right)}, 
\quad
\rho'(t,x)=
{\left(\frac{c}{t(1+\gamma t)}
-\frac{\ell (\ell-1)(\ell-2)\dots (\ell-2\ell) x_i x_i }{2 a (1+\ell d) {\left(t(1+\gamma t) \right)}^{2\ell+1}}
\right)}^{\ell d},
\ee
where $\gamma>0$, $c>0$ are constants and $t>0$.
In the latter equation it is assumed that $\ell$ is either integer or half-integer belonging 
to the sequence $\ell=\frac{1+4k}{2}$ with $k=0,1,2,\dots$, and $a$ is a constant 
which enters the equation of state.

%======================= Fig.4===========================>
\begin{figure}[ht]
\begin{center}
\resizebox{0.5\textwidth}{!}{%
\includegraphics{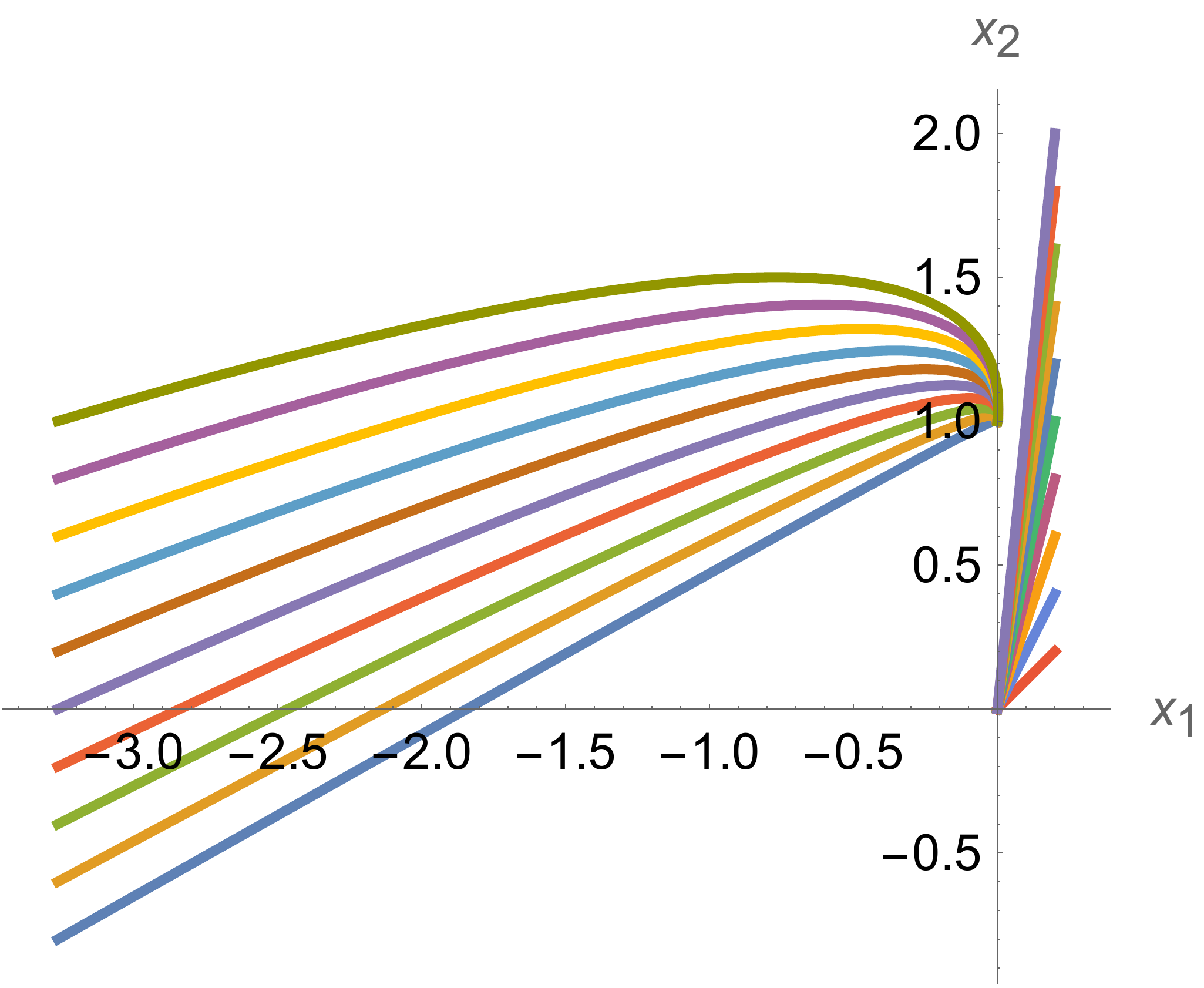}}\vskip-4mm
\caption{\small 
An example of a flow obtained by applying a specific acceleration transformation to the velocity vector 
field $\upsilon_i=\frac{\ell x_i}{t}$ for $\ell=1$ and $i=1,2$.
}
\label{fig4}
\end{center}
\end{figure}

In a similar fashion, one can use the higher order constant accelerations exposed in (\ref{tr}),
(\ref{trr1}), (\ref{trv1}) to generate novel solutions. Given a particular solution $\rho(t,x)$, $\upsilon_i (t,x)$
to the perfect fluid equations with the $\ell$-conformal Galilei symmetry, 
its deformation 
involving $2\ell+1$ constant vectors 
$\left(a^{(0)}_i,a^{(1)}_i,\dots,a^{(2\ell)}_i \right)$ reads
\be\label{SIS1}
\upsilon'_i (t,x)=\upsilon_i \left(t,x-\sum_{n=0}^{2\ell} a^{(n)} t^n \right)
+\sum_{n=0}^{2\ell} n a^{(n)}_i t^{n-1}, \qquad
\rho' (t,x)=\rho\left(t,x-\sum_{n=0}^{2\ell} a^{(n)} t^n \right).
\ee
In particular,
transformations of the latter type can be used to shape the velocity vector field entering (\ref{SIS})
in a desired way on a fixed time interval. For example, consider a bunch of rays in the first quadrant of a plane ($d=2$)
representing fluid particle orbits 
$x_i(t)=b_i t$ associated with the velocity 
vector field in (\ref{SIS}) for $\ell=1$, $i=1,2$ and the range of initial 
conditions $(b_1,b_2)=\left((0.1,0.1),(0.1,0.2),\dots,(0.1,1) \right)$.
Fig. 4 shows what happens to the flow if one takes the vector $a^{(0)}_i=(0,1)$, rotates it 
clockwise by the angle $\frac{2 \pi}{3}$ twice so as to 
generate the vectors $a^{(1)}_i=(\frac{\sqrt{3}}{2},-\frac 12)$ and $a^{(2)}_i=(-\frac{\sqrt{3}}{2},-\frac 12)$ and 
finally uses the triplet 
to construct $\upsilon'_i (t,x)$ in (\ref{SIS1}).\footnote{Solving the orbit equation
$\frac{d x_i(t)}{dt}=\upsilon_i (t, x(t))$ one gets 
$x_1 (t)=b_1 t-\frac{\sqrt{3} t^2}{2}$, $x_2 (t)=b_2 t+1-\frac{t^2}{2}$. In Fig. 4,
the range of initial 
conditions is chosen as before: 
$(b_1,b_2)=\left((0.1,0.1),(0.1,0.2),\dots,(0.1,1) \right)$.} Note that on the large time intervals 
the velocity vector field will tend to approach
the direction determined by the vector $a^{(2\ell)}_i$.

Finally, because the equations of motion (\ref{pfl}) are invariant under time 
translation, all the solutions above
can be modified so as to include an extra real parameter $t_0$ by the 
substitution $t \to t+t_0$.

\vspace{0.5cm}

\noindent
{\bf 4. Perfect fluid equations with the Lifshitz symmetry: Exact solutions}\\

The instance of $\ell=\frac 12$ in (\ref{algebra}) is known as the Schr\"odinger
algebra. If one disregards the generator of special conformal transformation $K$ 
in the Schr\"odinger algebra, the structure relations $[H,D]$ and $[D,C^{(1)}_i]$ 
can be modified so as to include an arbitrary constant $z$ known as the dynamical 
critical exponent (see e.g. \cite{MT}). The resulting algebra is referred to as the Lifshitz 
algebra\footnote{Similarly to the $\ell$-conformal Galilei algebra, here and in what follows
we ignore rotation generators.}
\be\label{LA}
[H,D]=z  H, \quad  [H,C^{(1)}_i]=C^{(0)}_i, \quad 
[D,C^{(0)}_i]=-\frac{1}{2} C^{(0)}_i, \quad
[D,C^{(1)}_i]=\left(z-\frac 12 \right) C^{(1)}_i,
\ee
where, as before, $i=1,\dots,d$. It is conveniently represented by the 
differential operators 
\be\label{LA1}
H=\frac{\partial}{\partial t}, \qquad D=z t \frac{\partial}{\partial t} 
+\frac{1}{2} x_i \frac{\partial}{\partial x_i}, \qquad 
C^{(0)}_i=\frac{\partial}{\partial x_i}, \qquad C^{(1)}_i=t \frac{\partial}{\partial x_i},
\ee
acting in a nonrelativistic spacetime parameterized by $(t,x_i)$.
In particular, $D$ in (\ref{LA1}) gives rise to the anisotropic scaling 
transformations
of the temporal and spatial coordinates
\be\label{act}
t'=e^{\lambda z} t, \qquad x'_i=e^{\frac{\lambda}{2}} x_i,
\ee
$\lambda$ being the transformation parameter. 

In order to guarantee that the perfect fluid equations hold invariant under the 
Lifshitz group,
one has to modify the equation of state accordingly \cite{AG1}
\be\label{L}
\frac{\partial \rho}{\partial t} + 
\frac{\partial ( \rho \upsilon_i )}{\partial x_i}=0,
\qquad \rho \mathcal{D} \upsilon_i =-\frac{\partial p}{\partial x_i}, \qquad
p=a \rho^{1+\frac{2(2z-1)}{d}},
\ee
where $\mathcal{D}=\frac{\partial}{\partial t} +
\upsilon_i  (t,x) \frac{\partial}{\partial x_i}$ is the material derivative and
$a$ is a positive constant. Transformation laws of the coordinates and fields 
under the Lifshitz group read \cite{AG1} (each transformation is separated by semicolon)
\begin{align}\label{TRL}
&
t'=t+\beta, && x'_i=x_i, &&
\nonumber\\[2pt]
&
\upsilon'_i (t',x')=\upsilon_i (t,x), && \rho'(t',x')=\rho(t,x);
\nonumber\\[6pt]
&
t'=e^{\lambda z}t, && x'_i=e^{\frac{\lambda}{2}}x_i, &&
\nonumber\\[2pt]
&
\upsilon'_i (t',x')=e^{-\lambda\left(z-\frac 12 \right)}\upsilon_i (t,x), && 
\rho' (t',x')= e^{-\frac{\lambda d}{2}}  \rho(t,x);
\nonumber\\[6pt]
&
t'=t, && x'_i=x_i+a^{(0)}_i,
\nonumber\\[2pt]
&
\upsilon'_i (t',x')=\upsilon_i (t,x), && \rho'(t',x')=\rho(t,x);
\nonumber\\[6pt]
&
t'=t, && x'_i=x_i+t a^{(1)}_i,
\nonumber\\[2pt]
&
\upsilon'_i (t',x')=\upsilon_i (t,x)+a^{(1)}_i, && \rho'(t',x')=\rho(t,x).
\end{align}

Let us generalize the analysis in the preceding section by
building a particular solution to (\ref{L}) which 
links to the anisotropic scaling transformations (\ref{act}). 

According to (\ref{TRL}), the generator of infinitesimal 
anisotropic scaling transformations reads
\be
D=z t \partial_t +\frac{1}{2} x_i \partial_i-\frac{d}{2} \rho \partial_\rho-(z-\frac 12) \upsilon_i \partial_{\upsilon_i}.
\ee
Analyzing the system of characteristic equations associated with the linear 
partial differential equation
$D f(t,x,\rho,\upsilon)=0$, one finds the invariant variables and fields
\be\label{supp4}
\frac{x_i}{t^{\frac{1}{2z}}}:=y_i, \qquad t^{\frac{d}{2z}} \rho(t,x):=w(y), \qquad
t^{1-\frac{1}{2z}} \upsilon_i (t,x):=u_i (y),
\ee
where $t>0$.

As above, the key observation is that, being rewritten in terms of the invariant objects, 
the continuity equation simplifies to the form
\be\label{CE3}
\frac{\partial}{\partial y_i} \left(w \left(u_i-\frac{y_i}{2z} \right) \right)=0. 
\ee
Instead of solving this partial differential equation in full generality, one can use it to fix
the velocity vector field
\be
u_i=\frac{y_i}{2z} \qquad \Rightarrow \qquad \upsilon_i (t,x)=\frac{x_i}{2zt}.
\ee
Then the Euler equation can be cast into the form
\be\label{EE3}
\frac{\partial}{\partial y_i} \left(w^{\frac{2(2z-1)}{d}} 
-\frac{{(2z-1)}^2 y_j y_j}{4a z^2 (d+2(2z-1))}  \right)=0 \quad \Rightarrow \quad
w(y)={\left(c+\frac{{(2z-1)}^2 y_j y_j}{4 a z^2 (d+2(2z-1))} \right)}^{\frac{d}{2(2z-1)}},
\nonumber 
\ee
where $c>0$ is a constant of integration. The latter relation allows one to determine the density 
via (\ref{supp4}). 

Thus, a particular solution to the perfect fluid equations with the Lifshitz symmetry (\ref{L}),
which links to the subgroup of anisotropic scaling transformations, reads
\be\label{LC}
\upsilon_i (t,x)=\frac{x_i}{2 z t}, \qquad \rho(t,x)=
{\left(\frac{c}{t^{2-\frac{1}{z}}}+\frac{{(2z-1)}^2 x_i x_i}{4az^2 (d+2(2z-1)) t^2} \right)}^{\frac{d}{2(2z-1)}}.
\ee
where $c>0$ is a constant and $a$ is a positive constant entering the equation of 
state in (\ref{L}). 

%======================= Fig.5===========================>
\begin{figure}[ht]
\begin{center}
\resizebox{0.5\textwidth}{!}{%
\includegraphics{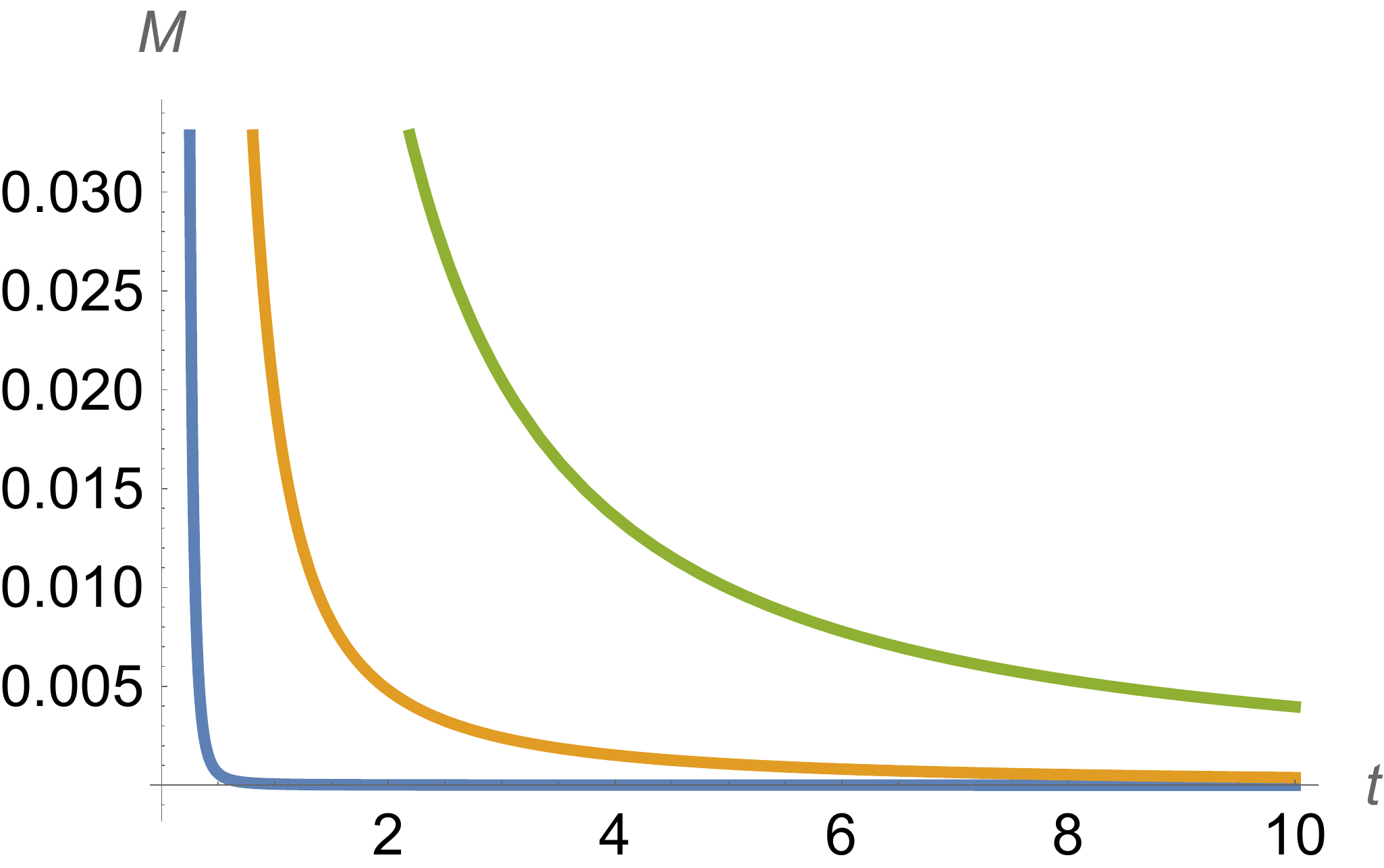}}\vskip-4mm
\caption{\small 
The dependence of the mass $M$ of a disk of unit radius centered at the origin of the coordinate system upon time $t$
for $z=0.6$ (blue), $z=0.7$ (orange), $z=0.8$ (green) in two spatial dimensions 
with $c=0.1$, $a=0.5$, and $t \in [0.1,10]$.
}
\label{fig5}
\end{center}
\end{figure}

A few comments are in order. Firstly, on physical grounds
the first term in braces should decrease over time, which
provides a natural lower bound on the dynamical 
critical exponent 
\be
z>\frac 12.
\ee
Interestingly enough, a similar bound has recently 
been revealed in \cite{AG4}, when building 
dynamical realizations of the Lifshitz group in mechanics and general relativity.
Secondly, 
the dependence of density upon $z$ in general correlates with
the values of the free parameters $c$ and $a$ chosen.
For example, Fig. 5 depicts
the dependence of the mass $M$ of a disk of unit radius centered at the origin of the coordinate system upon time $t$
for $z=0.6$ (blue), $z=0.7$ (orange), $z=0.8$ (green) in two spatial dimensions 
with $c=0.1$, $a=0.5$, and $t \in [0.1,10]$. Thirdly, 
the solution is isotropic. Finally, a particular 
solution (\ref{LC}) can be deformed so as to include extra free parameters by
making use of the temporal and spatial translations as well as the Galilei
boost in (\ref{TRL}).

\vspace{0.5cm}

\noindent
{\bf 5. A viscous fluid with 
the $\ell$-conformal Galilei symmetry}\\

In this section, we briefly discuss how the analysis in the preceding sections 
can be extended
to the case of a viscous fluid with 
the $\ell$-conformal Galilei symmetry.

In order to take into account the effects of viscosity, it suffices to introduce 
the rate-of-strain tensor 
(see e.g. in \cite{LL})
\be
\sigma_{ij}=\eta \left(\frac{\partial \upsilon_i}{\partial x_j} 
+\frac{\partial \upsilon_j}{\partial x_i}-\frac{2}{d} \delta_{ij} 
\frac{\partial \upsilon_k}{\partial x_k} \right)+\xi \delta_{ij}
\frac{\partial \upsilon_k}{\partial x_k},
\ee
which is the sum of the rate-of-shear (traceless) tensor and the rate-of-expansion tensor,
$\eta(t,x)$ and $\xi(t,x)$ being the shear and the volume viscosity coefficients, 
respectively, and then
include into the Euler equation 
the divergence of $\sigma_{ij}$
\be\label{vf}
\rho  \mathcal{D}^{2\ell} \upsilon_i=-\frac{\partial p}{\partial x_i}+ \frac{\partial \sigma_{ji}}{\partial x_j}.
\ee
It is assumed that the continuity equation and the equation of state in (\ref{pfl}) remain unchanged.

Taking into account (\ref{trr}), (\ref{trr1}), (\ref{trv}), (\ref{trv1}) and
analyzing the transformation of $\frac{\partial \sigma_{ji}}{\partial x_j}$ under the $\ell$-conformal Galilei
group, one concludes that (\ref{vf}) holds invariant under the acceleration transformations,
provided the shear and the volume viscosity coefficients transform as scalars.
As for the $SL(2,R)$ subgroup,
the temporal translation ($t'=t+a$, $x'_i=x_i$) and the dilatation ($t'=e^b t$, $x'_i=e^{b \ell} x_i$) are symmetries of
(\ref{vf}) provided the viscosity coefficients transform as follows
\be\label{trvf}
\eta(t,x)={\left(\frac{\partial t'}{\partial t} \right)}^{\ell d} \eta' (t',x'), \qquad 
\xi(t,x)={\left(\frac{\partial t'}{\partial t} \right)}^{\ell d} \xi' (t',x').
\ee
The invariance under
the special conformal transformation ($t'=\frac{t}{1-c t}$, 
$x'_i=\frac{1}{{(1-c t)}^{2\ell}} x_i$) is only feasible for the vanishing volume viscosity, 
$\xi(t,x)=0$, and $\eta(t,x)$ transforming as in (\ref{trvf}) (see also the discussion in \cite{RS,HZ}).

As far as exact solutions are concerned, the most interesting specimen originates from 
the subgroup of scaling transformations. Taking into account (\ref{trvf}) 
and the transformation law of the density 
in (\ref{trr}), one obtains a closed set of partial differential equations by
imposing the extra equations of state
\be
\eta(t,x)=\eta_0 \rho(t,x), \qquad \xi(t,x)=\xi_0 \rho(t,x), 
\ee
where $\eta_0$ and $\xi_0$ are positive constants.

Repeating the arguments in Sect. 3.2, one readily finds a 
solution for integer values of $\ell$
\be
\upsilon_i=\frac{\ell x_i}{t}, \qquad \rho(t,x)=\frac{c}{t^{\ell d}}, 
\qquad \eta(t,x)=\frac{c \eta_0}{t^{\ell d}}, \qquad \xi(t,x)=\frac{c \xi_0}{t^{\ell d}},
\ee
where $c>0$ is a constant. For half-integer $\ell$, the Euler equation results in
the transcendent equation to fix $w(y)$ in (\ref{vaf})
\be
a (\ell d+1) w(y)^{\frac{1}{\ell d}}-\xi_0 \ell d \ln{w(y)}
+\frac 12 \ell (\ell-1) \dots (\ell-2\ell)  y_i y_i=c, 
\ee
where $c$ is an arbitrary constant and $a$ is a constant entering (\ref{pfl}), 
which makes it difficult 
to find $\rho$, $\eta$, and $\xi$ explicitly.

A viscous fluid with the Lifshitz symmetry can be analyzed in a similar fashion.

\vspace{0.5cm}

\noindent
{\bf 6. Conclusion}\\

To summarize, in this work exact solutions to the perfect fluid equations with 
the $\ell$-conformal Galilei symmetry (\ref{pfl}) have been constructed within
the group-theoretic approach \cite{Ov,PO}. As the first step, the case of $\ell=\frac 12$
and one spatial dimension has been analyzed in full generality. 
Each one-dimensional subgroup 
in the full symmetry group was analyzed in turn and the corresponding exact solutions 
were built.
It was demonstrated that the most interesting specimen in the family
of exact solutions was associated with the subgroup of scaling transformations. In particular,
the scale-invariant variables and fields were found in terms of which
the continuity equation and the Euler equation reduced to 
ordinary differential equations. Two options were revealed to solve the latter.
Either one could use the continuity equation to fix the density and the Euler equation
to determine the velocity or vice versa. It was shown that the latter 
possibility, 
in which the continuity equation and the Euler equation effectively interchanged 
their roles, persisted to the case of arbitrary $\ell$ and 
arbitrary spatial dimension and allowed us to avoid direct solving of 
complicated partial differential equations.
Interestingly enough, the resulting 
velocity vector field proved to be a natural generalization of the Bjorken flow \cite{B} 
to arbitrary dimension and 
the group parameter $\ell$ naturally linked to the expansion rate. 
The analysis was 
also extended to similar perfect fluid equations
invariant under the Lifshitz group as well as to
the case of a viscous fluid with 
the $\ell$-conformal Galilei symmetry.

Let us discuss possible further developments. As was mentioned above,
by adjusting the value of $\ell$ and other free parameters available at hand
one can reach arbitrarily high density of a fluid (and hence
pressure) for a short period of time. It would be interesting to understand whether 
the fluid equations with the $\ell$-conformal Galilei symmetry or the Lifshitz symmetry 
may prove useful in other physical contexts such as the quark-gluon plasma, cosmology of the early universe 
and 
physics of explosion phenomena. 
An extension of the present analysis to the supersymmetric 
case \cite{AG2,TS4} is an interesting avenue to explore.

\vspace{0.5cm}

\noindent{\bf Acknowledgements}\\

\noindent
This work was supported by the Russian Science Foundation, grant No 23-11-00002-Ext.


\begin{thebibliography}{nn}
\bibitem{MR}
M. Rangamani, {\it Gravity and hydrodynamics: Lectures on the fluid-gravity correspondence}, 
Class. Quant. Grav. {\bf 26} (2009) 224003, arXiv:0905.4352.
\bibitem{RS}
L. O'Raifeartaigh, V.V. Sreedhar, {\it The maximal kinematical invariance group of fluid
dynamics and explosion-implosion duality}, Annals Phys. {\bf 293} (2001) 215, hep-th/0007199.
\bibitem{Ov}
L.V. Ovsiannikov, {\it Group Analysis of Differential Equations}, Academic Press, 1982.
\bibitem{FO}
I. Fouxon, Y. Oz, {\it CFT hydrodynamics: symmetries, exact solutions and gravity}, JHEP
{\bf 0903} (2009) 120, arXiv:0812.1266.
\bibitem{Henkel}
M. Henkel, {\it Local scale invariance and strongly anisotropic equilibrium critical systems}, Phys. Rev. Lett. {\bf 78} (1997) 1940, cond-mat/9610174.
\bibitem{NOR}
J. Negro, M.A. del Olmo, A. Rodriguez-Marco, {\it Nonrelativistic conformal groups}, J. Math. Phys. {\bf 38} (1997) 3786.
\bibitem{AG}
A. Galajinsky, {\it Equations of fluid dynamics with the 
$\ell$-conformal Galilei symmetry}, Nucl. Phys. B {\bf 984} (2022) 115965, arXiv:2205.12576.
\bibitem{AG1}
A. Galajinsky, {\it Group-theoretic approach to perfect fluid equations with 
conformal symmetry}, Phys. Rev. D {\bf 107} (2023) 026008, arXiv:2210.14544.
\bibitem{TS}
T. Snegirev, {\it Hamiltonian formulation for perfect fluid equations with the $\ell$-conformal
Galilei symmetry}, J. Geom. Phys. {\bf 192} (2023) 104930, arXiv:2302.01565.
\bibitem{TS1}
T. Snegirev, {\it Perfect fluid coupled to a solenoidal field which enjoys the $\ell$-conformal
Galilei symmetry}, Nucl. Phys. B {\bf 1002} (2024) 116526, arXiv:2312.10507.
\bibitem{TS2}
T. Snegirev, {\it Lagrangian formulation for perfect fluid equations with the $\ell$-conformal
Galilei symmetry}, Phys. Rev. D {\bf 110} (2024) 045003, arXiv:2406.02952.
\bibitem{TS3}
T. Snegirev, {\it Perfect fluid dynamics with conformal Newton-Hooke symmetries},
Nucl. Phys. B {\bf 1015} (2025) 116902, arXiv:2501.16781.
\bibitem{BJLNP}
B. Bistrovic, R. Jackiw, H. Li, V.P. Nair, S.-Y. Pi, {\it Non-abelian fluid dynamics in Lagrangian formulation}, 
Phys. Rev. D {\bf 67} (2003) 025013, hep-th/0210143.
\bibitem{NRR}
V.P. Nair, R. Ray, S. Roy, {\it Fluids, anomalies and the chiral magnetic effect: A group-theoretic formulation}, 
Phys. Rev. D {\bf 86} (2012) 025012, arXiv:1112.4022.
\bibitem{GM4}
A. Galajinsky, I. Masterov, {\it On dynamical realizations of $\ell$-conformal Galilei and Newton-Hooke algebras}, Nucl. Phys. B {\bf 896} (2015) 244, arXiv:1503.08633.
\bibitem{CG}
D. Chernyavsky, A. Galajinsky, {\it Ricci-flat spacetimes with $\ell$-conformal Galilei symmetry}, Phys. Lett. B {\bf 754} (2016) 249, arXiv:1512.06226.
\bibitem{HH2}
M. Hassaine and P.A. Horvathy, {\it Field dependent symmetries of a nonrelativistic fluid model},
Annals Phys. {\bf 282} (2000) 218, math-ph/9904022.
\bibitem{JNPP}
R. Jackiw, V.P. Nair, S.Y. Pi, A.P. Polychronakos, {\it Perfect fluid theory and its extensions}, J. Phys. A {\bf 37} (2004) R327, arXiv:hep-ph/0407101.
\bibitem{HH1}
M. Hassaine, P.A. Horvathy, {\it Symmetries of fluid dynamics with
polytropic exponent}, Phys. Lett. A {\bf 279} (2001) 215, hep-th/0009092.
\bibitem{BMW}
S. Bhattacharyya, S. Minwalla, S.R. Wadia, {\it The incompressible non-relativistic Navier-Stokes equation from gravity}, JHEP {\bf 0908} (2009) 059, arXiv:0810.1545.
\bibitem{HZ}
P.A. Horvathy, P.-M. Zhang, {\it Non-relativistic conformal symmetries in fluid mechanics}, Eur. Phys. J. C {\bf 65} (2010) 607, arXiv:0906.3594.
\bibitem{PO}
P.J. Olver, {\it Applications of Lie Groups to Differential Equations}, Graduate Texts in Mathematics,
Springer-Verlag New York Inc., 1986.
\bibitem{MT}
M. Taylor, {\it Lifshitz holography}, Class. Quant. Grav. {\bf 33} (2016) 033001, arXiv:1512.03554.
\bibitem{B}
J.D. Bjorken, {\it Highly relativistic nucleus-nucleus collisions: The central rapidity region}, Phys. Rev. D {\bf 27} (1983) 140.
\bibitem{AG4}
A. Galajinsky, {\it Dynamical realizations of the Lifshitz group}, Phys. Rev. D {\bf 105} (2022) 106023, arXiv:2201.10187.
\bibitem{AG3} 
A. Galajinsky, {\it Remarks on higher Schwarzians}, Phys. Lett. B {\bf 843}
(2023) 138042, 2302.00317.
\bibitem{SM}
V.I. Smirnov, {\it A Course of Higher Mathematics}, Vol. 2, Pergamon Press, 1964.
\bibitem{LL}
L.D. Landau, E.M. Lifshitz, {\it Fluid Mechanics}, Pergamon Press, 1959.
\bibitem{AG2} 
A. Galajinsky, {\it Equations of fluid mechanics with $N=1$ Schrodinger supersymmetry},
Nucl. Phys. B {\bf 999} (2024) 116450, arXiv:2312.04084.
\bibitem{TS4}
T. Snegirev, {\it Perfect fluid equations with $N=1,2$ Schrodinger supersymmetry}, Mod. Phys. Lett. A {\bf 41} (2026) 2550214,
arXiv:2505.22043.
\end{thebibliography}
\end{document}